# First principles study of optical and tunable electronic properties of crystalline Li$_2$TeO$_3$

Aditya Dey

*Department of Physics, Indian Institute of Technology Patna, Bihta Campus, Patna, Bihar, India-801106*

**Abstract -** The optical and electronic properties of crystalline Li$_2$TeO$_3$, which is a tellurite glass, is studied in the framework of density functional theory (DFT) implemented software SIESTA. The material has monoclinic symmetrized structure and the unit or primitive cell of the material, periodic in all directions has been taken to study the properties. The electronic structures show that it is a wide-gap semiconductor and the property changes to metallic when subjected to electric field. This tunable property can be used in various fields of electronics. The optical properties studied tells that Li$_2$TeO$_3$ can be a promising material to be used as a hole transport material (HTM) for developing efficient perovskite solar cell including other applications as well.

The studies about the tellurite glasses show that it is worthy of scientific interest because of its physical and optical properties like low melting points, high dielectric constant, higher refractive index and good infrared transmissivity [1-3]. Also, among the known glasses, it has the largest non-linear susceptibility enabling it to be used as optical switching devices [4]. Furthermore, tellurite glasses possess chemical durability and are water resistive [5]. The functional properties of TeO$_2$ glasses as an active material in optical amplifiers have been studied by the insertion of modifiers such as alkali oxides or transition metal oxides which are added to enhance the ease of glass formation [6-7]. Compared with silicate and phosphate glasses, tellurites exhibit relatively higher refractive indices, dielectric constants and lower phonon energies [8]. Although the previous works show that the tellurite glasses have attractive physical and optical properties, but many of these materials have not extensively studied yet. In this study, one of these materials

$Li_2TeO_3$ is considered and its electronic structure and optical properties has been investigated. The crystalline $Li_2TeO_3$ has been studied which contains the $TeO_3$ pyramidal structure, which is believed to be present in the doped glasses as well. Some of these reports include the theoretical Raman and IR spectrum of $Li_2TeO_3$, computed within density functional linear response theory and vibrational properties as well [9-10]. To dig deeper into the properties of this material, this study has been performed which led to interesting results and simultaneously opened up possibilities of using this material in developing perovskite solar cells and can have other electronic applications as well.

The calculations are done using ab-initio density functional theory (DFT) as implemented in the software SIESTA [11]. A real space mesh cutoff of 300 Ry has been used and the generalized-gradient approximation (GGA) with the Perdew-Burke-Ernzerhof (PBE) form is chosen for the exchange-correlation functional [12]. Normconserving pseudopotentials in the fully nonlocal Kleinman− Bylander form have been considered for all the atoms [13]. Conjugate-gradient (CG) method [14] has been used to optimize the structures and a double-ζ polarized (DZP) basis set is used. Systems are considered to be relaxed only when the forces acting on all the atoms are less than 0.01 eV Å$^{-1}$. Firstly, the unit cell of the material is taken and optimization is done proceeding with the calculation of electronic and optical properties. A 12 X 12 X 12 k-point grid has been considered within the Monkhorst−Pack scheme [15] for the optimization and calculation of these properties. Further, electric field is applied along *c* axis (z-direction) to study the modulation of electronic structure. For this, a vacuum of 24 Å is taken in the *c* axis and the k-point grid considered is 24 X 24 X 1. For optical properties, an optical mesh of 20 X 20 X 20 is taken with an optical broadening of 0.1 eV.

The structure of the crystalline $Li_2TeO_3$ is shown in Figure 1. It has a face centred monoclinic symmetry with *C*2/*c* space group having four formula units per unit cell. Taking the experimental parameters, optimization is done and the relaxed cell parameters are a = 5.09, b = 9.597 and c = 13.762 Å, which are almost equal to the experimental values [16]. The structure is formed by bilayers of $TeO_3$ pyramidal units which are coordinated by Li cations which form a layer lattice consisting of distorted $LiO_4$ tetrahedra and $TeO_3$ pyramids.

After taking the optimized geometry of the material, the electronic structures are calculated. This includes the band structure, density of states (DOS) and partial density of states (PDOS). Firstly, these calculations are done in the absence of external electric field ($E_{ext}$) and then $E_{ext}$ is applied in the out of the plane direction which is along the *c* axis (z-direction) and perpendicular to xy plane. Figure 2 illustrates the electronic structures for both the cases. The brillouin zone taken is that of a face centred monoclinic lattice and the high symmetry points are marked in the band structure which shows that the pristine $Li_2TeO_3$ ($E_{ext} = 0$) is an indirect wide gap semiconductor with a band gap ($E_g$) of 3.71 eV. The conduction band minimum is attained at X point and the valence band maximum at $\Gamma$ point. The DOS plot also indicates the same as there are no states near the Fermi level which shows an energy gap between the conduction and valence bands (CB and VB). For a better understanding, PDOS is plotted for the outermost orbitals of the atoms whose configurations are $2s^1$ for Li, $5p^4$ for Te and $2p^4$ for O as these are mainly responsible for the states in the CB and VB (Figure 2). For $E_{ext} = 0$, the PDOS plot shows that the major peaks in the VB at -2.5 eV is predominantly due to the O 2p states and the peaks in CB occurs with the contribution of both O 2p and Te 5p states. The contribution of Li 2s states is almost negligible. The O 2p states play an important role in the attained band gap.

Next, electric field strength varying between 10-50 V/nm is applied and on its application, it is observed from Table 1 that the band gap drastically reduces on applying the external field and continues to reduce with increasing magnitude of $E_{ext}$ and at 30 V/nm $E_g$ becomes zero suggesting that the material shows metallic behavior undergoing a property transition. Figure 2 shows the electronic structures with $E_{ext} = 30$ V/nm. Comparing the band structures with and without electric field, it can be observed that on application of $E_{ext}$, the CB and VB move towards the Fermi level and crosses it which induces the metallic nature of $Li_2TeO_3$. The DOS and PDOS plot for $E_{ext} = 30$ V/nm also explains this by showing that there are states present in the vicinity of the Fermi level. The metallic nature is mainly because of the O 2p states with contribution from Te 5p states and marginally from Li 2s states as well as observed from the PDOS plot. The contribution of states for the peaks in CB and VB is similar to what observed for without $E_{ext}$. The reduction in band gap and the property transition by applying electric field is mainly due to the shift in band energies caused by subtle movement of the electron density because of $E_{ext}$. Increasing the field strength triggers the electron density movement and shift in band energies also causing shifting of energy states. The PDOS plot further explains that with

$E_{ext}$, the states have moved towards the Fermi level with the major movement of electron density of O 2p states. Hence, the deduced electronic structures of $Li_2TeO_3$ with and without $E_{ext}$ show an interesting behavior that it can be tuned to be used both as a semiconductor or a metal.

The optical properties of $Li_2TeO_3$ have been calculated for polarized light along both in plane ($\mathbf{E}\perp c$) and out of the plane ($\mathbf{E}\|c$) directions of light and the properties which are calculated are real ($\varepsilon_1$) & imaginary ($\varepsilon_2$) parts of dielectric function, absorption coefficient ($\alpha$), reflectance ($R$), extinction coefficient ($k$) and refractive index ($n$) in the energy range 0 - 30eV. It is observed that the obtained properties are not anisotropic (same properties in both directions of polarized light) and does not depend on the direction of polarized light. So, the properties for one of the directions, out of the plane ($\mathbf{E}\|c$) polarization are discussed. All the calculated optical properties are shown in Figure 3.

The real and imaginary parts of the dielectric function is obtained from the expression $\varepsilon(\omega) = \varepsilon_1(\omega) + i\varepsilon_2(\omega)$, where $\varepsilon(\omega)$ is the complex dielectric function. $\varepsilon_1$ explains about the electronic polarizability and anomalous dispersion of the material and $\varepsilon_2$ is associated with dissipation of energy into the medium. The peaks of $\varepsilon_1$ occur in the near UV region at 4.01 and 4.48 eV and the static dielectric constant (at zero energy) is 2.90. For $\varepsilon_2$, the maximum peak occurs at 4.76 eV in the near UV region and other peaks are also observed in ~7-10 eV that lies in far UV region. However, $\varepsilon_2$ is zero in the visible region that explains no absorbance in that region. This is also observed in the absorption spectra which show that value of absorption coefficient ($\alpha$) of the material is zero up to 3.79 eV including the visible region thus indicating that without any attenuation, the visible light can pass almost easily through this material. The highest peaks of $\alpha$ is observed in the UV region suggesting that there are low electron losses in this range. The reflectance curve of the material shows that it has a maximum reflectivity of only 20% and the major peaks lie in the UV range only. In the visible region, the reflectance rises from 8% to 12% which explains that the material will not reflect much of the incident visible light. The spectrum for extinction coefficient ($k$) shows that the $k$ value is also zero up to 3.80 eV, similar to the value of $\alpha$, again suggesting that it will not attenuate most of the incident visible light. The prominent peaks of $k$ lie in the UV range, which tells that the penetration depth of the incident photons would be the least at this energy. The static refractive index $n$ of $Li_2TeO_3$ is 1.71 and attains a highest value of 2.40 at 4.45 eV in the UV region. The *n* value rises in the visible region from 1.74 to 2.02.

The optical properties arise mainly because of the interband transition which occurs by electronic transition of O 2p bands as seen in the PDOS (Figure 2). The obtained properties of $Li_2TeO_3$ in the visible region show that it can have an interesting application as a HTM layer which is mainly used in the perovskite solar cells. HTMs play an important role in extracting and transporting photogenerated holes, thus minimizing the recombination losses in perovskite solar cells and thus achieving high performances [17-18]. Thus they ought to possess stable optical properties with suitable light absorption in visible and near-IR region of the solar spectrum. The optical properties of $Li_2TeO_3$ shows that using this as a HTM layer, it will not enervate much of the incident visible solar light and contribute towards the efficiency of solar cell. This is supported by low reflectivity and zero *α* and *k* values in the visible region and also the value of *n* which aligns well with the refractive indices of other HTMs [19-20]. Apart from this, $Li_2TeO_3$ can be used as UV absorbers to protect sensitive devices from exposure to UV radiations.

Hence, in this article, calculations are performed through first principles calculations using DFT to study the electronic and optical properties of a crystalline material $Li_2TeO_3$ which is a kind of tellurite glass. The structural parameters showed that it has a face centered monoclinic lattice. The electronic structure showed that it is a wide band gap semiconductor with an energy gap of 3.71 eV, which suggests it can have plenty of semiconducting applications. Further, application of electric field along *c* axis caused reduction in band gap and beyond 40 V/nm the material showed metallic behavior from which one can say that the electronic property is easily tunable. The PDOS plot reveals that O 2p states are majorly contributing in the semiconducting and metallic behavior. Next, the optical properties in 0-30 eV range are calculated. The observation of these properties in the visible region and the attained wide band gap of the material show that it will not attenuate or absorb most of the visible light incident on it. These properties tell that $Li_2TeO_3$ has a great potential to be used as a hole transport material mainly used in the perovskite solar cell and can help in increasing the efficiency of the solar cell. Therefore, these findings including the optical properties and tunable electronic properties apart from having potential use in solar cells can have many other applications in fields like optoelectronics.

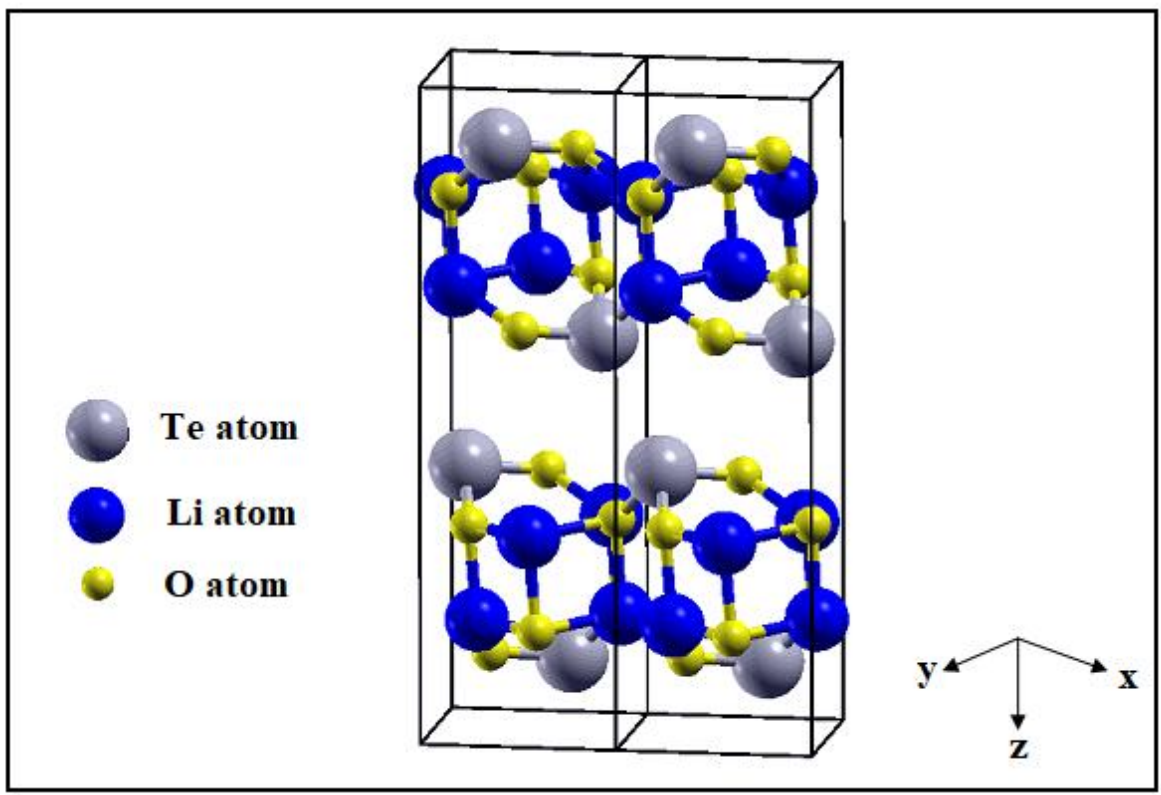

**Figure 1**

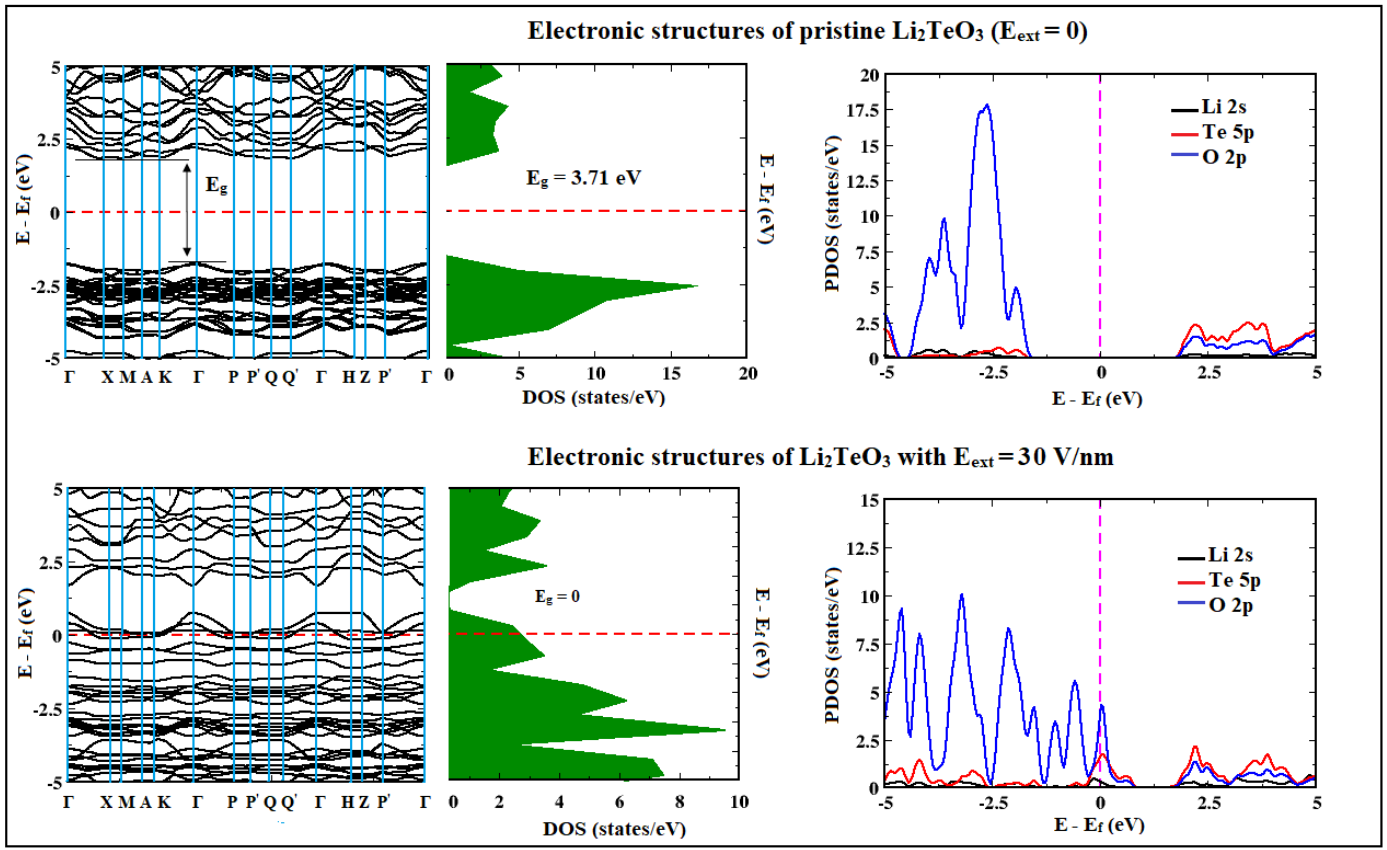

**Figure 2**

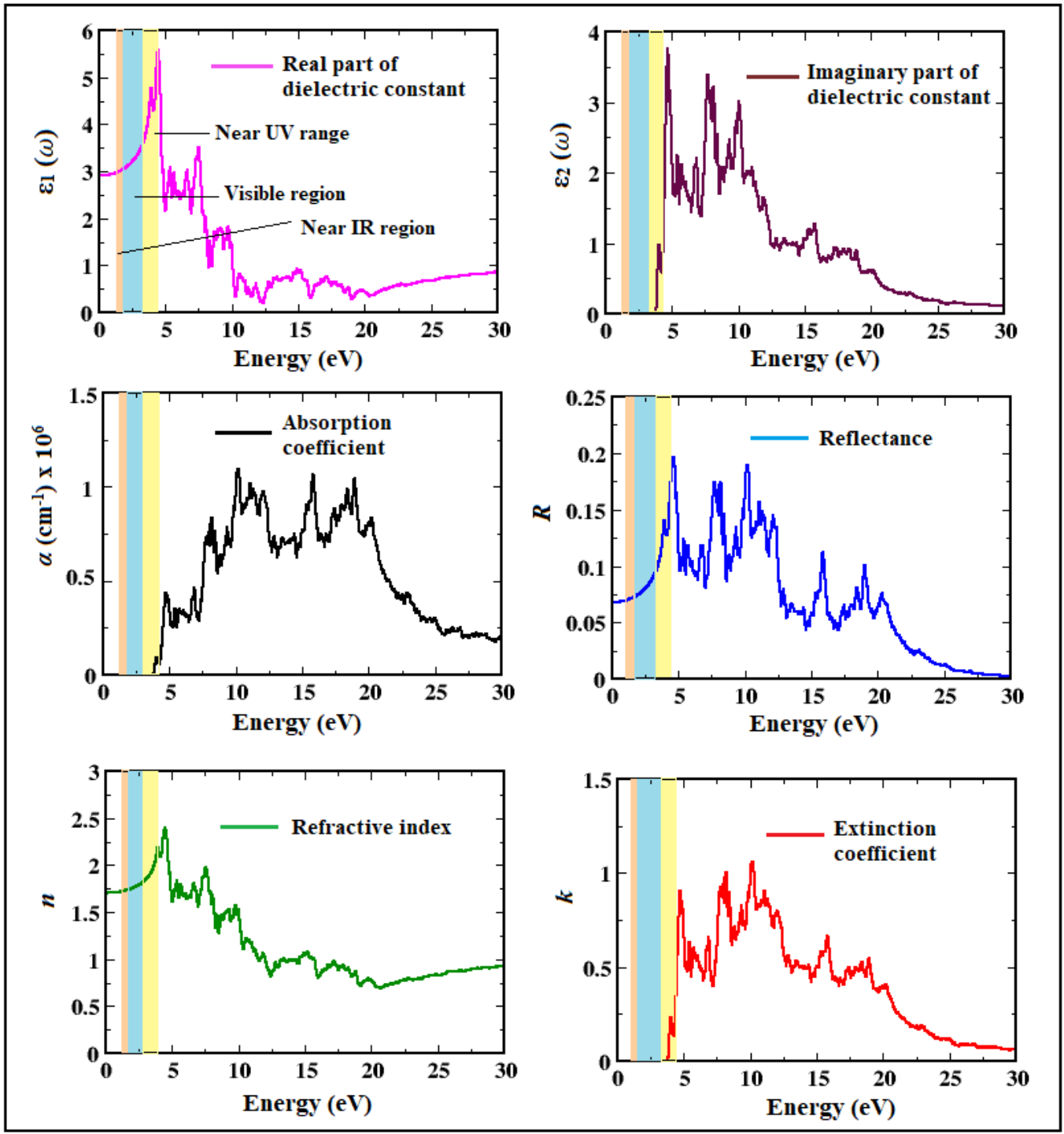

**Figure 3**

**Figure Captions**

**Figure 1** – Optimized structure of crystalline $Li_2TeO_3$. Boxes with black solid line represent the unit cell.

**Figure 2** – Electronic structures of $Li_2TeO_3$ with $E_{ext} = 0$ and $E_{ext} = 30$ V/nm. Band structure and DOS is shown on the left hand side and PDOS on right hand side.

**Figure 3** – Optical properties (*real & imaginary parts of dielectric constant, absorption coefficient, reflectance, refractive index and extinction coefficient*) of $Li_2TeO_3$. Different regions of electromagnetic spectrum are also illustrated.

| Electric field (V/nm) | Band gap, $E_g$ (eV) |
|---|---|
| 0 | 3.71 |
| 10 | 1.02 |
| 20 | 0.11 |
| 30 | 0.00 |
| 40 | 0.00 |
| 50 | 0.00 |

**Table 1** – Variation of band gap, $E_g$ with externally applied electric field $E_{ext}$